\documentclass[preprint,showpacs,superscriptaddress,aps]{revtex4-1}
\usepackage{amssymb}
\usepackage{amsmath,amssymb,graphicx}
\usepackage{hyperref}
\usepackage{graphicx}% Include figure files
\usepackage{dcolumn}% Align table columns on decimal point
\usepackage{bm}% bold math
\usepackage{color}
\def\be{\begin{equation}}
\def\ee{\end{equation}}
\def\bea{\begin{eqnarray}}
\def\eea{\end{eqnarray}}

\def\lsim{\raise0.3ex\hbox{$\;<$\kern-0.75em\raise-1.1ex\hbox{$\sim\;$}}}
\def\gsim{\raise0.3ex\hbox{$\;>$\kern-0.75em\raise-1.1ex\hbox{$\sim\;$}}}

\usepackage{pstricks}
\begin{document}

\title{New Physics effects in $D^+ \rightarrow K^- \pi^+  \pi^+ $}

\author{David Delepine}
\email{delepine@fisica.ugto.mx}
\affiliation{{\fontsize{10}{10}\selectfont{Division de Ciencias e Ingenier\'ias,
 Universidad de Guanajuato, C.P. 37150, Le\'on, Guanajuato, M\'exico.}}}

\author{Gaber Faisel}
\email{gfaisel@hep1.phys.ntu.edu.tw}
\affiliation{{\fontsize{10}{10}\selectfont{Department of Physics,
National Taiwan University, Taipei, Tawian 10617.}}}
\affiliation{{\fontsize{10}{10}\selectfont{Egyptian Center for
Theoretical Physics, Modern University for Information and
Technology, Cairo, Egypt.}}}

\author{ Carlos A. Ramirez}
\email{jpjdramirez@yahoo.com}
\affiliation{{\fontsize{10}{10}\selectfont{Depto. de F\'isica,
Universidad de los Andes, A. A. 4976-12340, Bogot\'a, Colombia.}}}

\begin{center}

\begin{abstract}
In this paper, we study the Cabibbo favored three body
non-leptonic $D^+ \rightarrow K^- \pi^+  \pi^+ $ decay. We show
that the corresponding direct CP asymmetry is so tiny in the
framework of the Standard Model and is out of the experimental
range. Motivated by this result we extend the study of the CP
asymmetry to include a toy model with CP violating weak phase
equals $20^{\circ}$ in $a_2$, a model with extra gauge bosons
within Left-Right Grand Unification models and a model with
charged Higgs boson. We show that the toy model can strongly
improve the SM prediction of the CP asymmetry to be about $30\%$.
The largest CP asymmetry can be achieved in the non-manifest
Left-Right models where a CP asymmetry up to $25\%$ can be
reached. For the two Higgs doublets models  the CP asymmetry is of
order $10^{-3}$.
\end{abstract}
\end{center}
\pacs{}

\maketitle
\section{Introduction}

The Standard Model of the strong and electroweak interactions (SM)
is one of the greatest scientific success of all the time. Until
now, all experimental data are compatibles with SM predictions and
the recent discovery of the Higgs particles has permitted to
complete the spectrum of particles as expected within the SM with
three fermion families. Also until now, the LHC has  disregarded
many extensions of the SM. But there are several hints that New
Physics should not be far from the corner. In particular, the
cosmological observation of the baryon asymmetry of the Universe
is clearly a hint for New Physics as the CP asymmetry in the SM is
not enough to explain it. Also Dark Matter introduced to explain
the rotational curves of the galaxies doesn't find its place
within SM.

 Within the Standard Model, CP violation in the quark sector is only produced through
 the phase which appears in the Cabibbo-Kobayashi-Maskawa quark mixing matrix \cite{Cabibbo:1963yz,Kobayashi:1973fv}.
 All the experimental results can be explained through the fitting of this unique phase\cite{ckmfitter}.
 A better accuracy in the measurement of CP asymmetries should permit to disentangle the CKM CP violating phase
 from New Physics CP violating sources. With LHCb and the new generation of B factories as BELLE II,
 the D and B physics will reach a degree of accuracy never seen until now and should be compelling
 in the search of New Physics with direct search for new particles done in LHC or ILC colliders.

It is why it is very interesting to look for CP asymmetries in
processes  forbidden or very suppressed within SM. The D physics
is offering us a new window for this search. Many channels of D
mesons decays have no or very suppressed CP asymmetries within SM.
In particular the Cabibbo Favored (CF)  D decays usually have no
SM phases at tree level and can be generated only at one loop
level within SM. This means that these channels can be seen as
smoking gun for the search for New Physics CP violation. In
previous work, we work on the CF non leptonic two body D decays
into pions and Kaons and we show that  within extensions with
extra Higgs or with non-manifest left-right symmetry large CP
asymmetries could be reached \cite{Delepine:2012xw}.

In this paper, we study the decay $D^+ \to K^-\pi^+\pi^+ $ which
is CF non leptonic D decay channel \cite{charmnl} with a large
branching ratio of 9.22(17) \% \cite{pdg}.  But as Cabibbo Favored
processes, it is expected not to have a significant CP violation
within Standard Model. So, it is an interesting channel to look
for new sources of CP violation.

In a first step, we quickly describe the way to parametrize the
 form factors (FF) of this decay mode using the FF models
developed in refs. \cite{Boito:2009qd,ElBennich:2009da}. It is
important to notice that any other modelisation of the form
factors does not change significantly the results presented in
this paper. Then, we check that the SM CP asymmetries is very
suppressed. In the third section, we introduce new CP violating
sources. First, we propose a model-independent approach, adding a
weak phase to $a_2$ contribution to this process and looking in
which phase space range one could expect a maximal CP asymmetries.
Then, we study in details two typical models of New Physics: one
assuming  an extra charged gauge boson and another model based on
non-manifest left right symmetry. We show that in both case, a
significant enhancement in the CP asymmetries is expected in some
phase space ranges compared to SM.

\section{ General Description of CF non leptonic $D^+\to K^-\pi^+\pi^+$ \label{gen}}
The effective Hamiltonian describing  $D^+\to K^-\pi^+\pi^+$  can
be written as
\begin{eqnarray}
{\cal H}_{\rm eff.} &=& {G_F\over\sqrt{2}}V_{cs}^*V_{ud}
\left[\sum_{i,\ a} c_{1ab}^i \bar s\Gamma^ic_a\bar u \Gamma_id_b+
\sum_{i,\ a} c_{2ab}^i \bar u\Gamma^ic_a\bar s \Gamma_id_b \right]
\end{eqnarray}
here  $i$ runs over $S,V$ and $T$ which stand for  scalar (S),
vectorial (V) and tensorial (T) operators respectively. The Latin
indexes $a,\ b=L,\ R$ and $q_{L,\ R}=(1\mp \gamma_5)q$.

In the SM, ${\cal H}_{\rm eff.}$ has only two operators \cite{charmnl} and  the
other operators can  be generated only in the presence of  new
physics. Thus we can write
\begin{eqnarray}
{\cal H}^{SM}_{\rm eff.}  &=&
{G_F\over\sqrt{2}}V_{cs}^*V_{ud}\left(c_1\bar s \gamma_\mu c_L\bar
u \gamma^\mu d_L+c_2\bar u \gamma_\mu c_L\bar s \gamma^\mu
d_L\right)+{\rm h.c.}  \label{SMH}
\end{eqnarray}

 The amplitude of the decay process $D^+(p)\to K^-(p_K)\pi^+(p_1)\pi^+(p_2)$ can be obtained via
\begin{eqnarray}
{\cal M} = \, <K^-\pi^+\pi^+ | {\cal H}^{SM}_{\rm eff.}|D^+> \,=
{G_F\over\sqrt{2}}V_{cs}^*V_{ud}\,{\cal A}
\end{eqnarray}
 For detailed discussion of the different contributions to
the amplitude from scalar, vector and so on we refer to
ref.\cite{Boito:2009qd}. The expression for ${\cal A}$ is given as
\cite{Boito:2009qd}

\begin{eqnarray} {\cal A} \equiv {\cal A}_s(s) + {\cal A}_p(s)+(s
\leftrightarrow t) \label{amplitude}
\end{eqnarray}

where $s$ and $t$ are the Mandelstam variables and the expressions of $A_s(s)$ and $A_p(s)$ are given as

\begin{eqnarray}
A_s(s) &=& \left[a_1f_\pi\chi_S^{\rm
eff.}(m_D^2-s)+a_2{\Delta_{D\pi}^2\Delta_{K\pi}^2\over
s}F_0^{D\pi}(s)\right]F_0^{K\pi}(s)
\nonumber \\
A_p(s) &=& -4\left[a_1f_\pi\chi_V^{\rm
eff.}+a_2F_+^{D\pi}(s)\right]F_+^{K\pi}(s)|{\bf p_K'}| |{\bf
p_2'}|\cos\theta'
\end{eqnarray}

where $a_1\equiv c_1+ c_2/N_c =1.2\pm 0.1$, $a_2\equiv
c_2-c_1/N_C=-0.5\pm 0.1$\cite{charmnl} and $N_C=3$ is
the color number. $\Delta_{D\pi}^2=m^2_D-m^2_{\pi}$,
$\Delta_{K\pi}^2=m^2_K-m^2_{\pi}$ and  $\theta'$ is the angle
between the direction of the momentum  of $K$ and the direction of
the momentum  $\pi_2$ in the $K\,-\,\pi_1$ center of mass frame.
The  Mandelstam variables are defined  from the Kinematic of the
process  as

\begin{eqnarray}
s &=& (p_K+p_1)^2,\ t=(p_K+p_2)^2,\ u=(p_1+p_2)^2,\
s+t+u=m_D^2+m_K^2+2m_\pi^2
\end{eqnarray}

In our analysis we use \cite{Boito:2009qd}

\begin{eqnarray}
F_{+,\ 0}^{D\pi}(x) &=& {F_+^{D\pi}(0) \over 1-x/m_{+,\
0}^2},\,\,\,\,\,\,\,\,\,\,\,\,\, F_+^{D\pi}(0)=F_0^{D\pi}(0)\simeq
0.624,
\end{eqnarray}
 where

 \begin{eqnarray}
 m_+=m_{D^{*0}}=2007\
{\rm MeV},\ m_0=m_0^{*0}=2352\ {\rm MeV}  \label{a2e}
\end{eqnarray}

\begin{eqnarray}
\chi_S^{\rm eff.} &\simeq & { g_{K_0^*K\pi }F_0^{DK_0^*}(m_\pi^2)
\over \Gamma_{K_0^*}(m_{K_0^*}^2)|F_0^{K\pi}(m_{K_0^*}^2 )|
}=4.4(28),\ 4.9(4)\ {\rm GeV}^{-1},\nonumber\\
\chi_V^{\rm eff.}&=&{g^*\sqrt{s}\ A_0\over m_*^2} \simeq
{g^*A_0\over m_*}=4.9(2),\ 4.4(6)\ {\rm GeV}^{-1}
\end{eqnarray}

with $F_0^{DK_0^*}(m_\pi^2)=1.24(7)$ \cite{Cheng:2002ai} and
$F_+^{DK^*}(m_\pi^2)=0.76(7)$
\cite{Melikhov:2000yu,Fajfer:2005ug}. For later analysis we list
the expressions for  the fit and interference fit partial
fractions which are defined as (see
\href{http://pdg.lbl.gov/reviews/rppref/mini/2012/dalitz_analysis_s031dpf-web.pdf}{Dalitz
plot analysis formalism} in \cite{pdg})

\begin{eqnarray}
f_i &=& {\int {\rm d}s{\rm d}u |A_i|^2 \over \int {\rm d}s{\rm d}u |A|^2 } \nonumber \\
f_{ij} &=& {2\int {\rm d}s{\rm d}u\ {\rm Re}\ (A_iA_j^*) \over
\int {\rm d}s{\rm d}u |A|^2 }
\end{eqnarray}

In Table\ref{fractions} we show the results of these partial
fraction in the case of the SM. The scalar and vectorial $\pi-K$
form factors were taken from refs.\cite{Boito:2009qd,
ElBennich:2009da}. In the rest of the paper we will discuss the
direct CP asymmetry within SM framework and some possible
extensions of the SM.

\begin{table}\centering
\begin{tabular}{|r|r|r|r|r|r|r|r|r|r|r|r|r|}   \hline
Model-fract.    &$\chi_S$& $\chi_P$ & $\phi_W$  & $f_i:\ S$ & $P$
&SPI&$2\pi^+$ &I & II & III & IV  \\ \hline PDG\cite{pdg}   & -&-
& -  &80.2(27)&11.1(3) &-   &15.4(5) &-&-&-&  \\ \hline
Escri.-Mouss.\cite{Boito:2009qd, ElBennich:2009da}  &4.99 &5.62
&0   &80.2 &16.4    &3.4 &0       &5 &23.6 &64.8 &6.6     \\
\hline $f_i$-Cleo fit  &4.99 &5.52 &0 &82.2 &14.9    &2.9 &0
&6   &22.2 &62.8 &9.1    \\  \hline
\end{tabular}
\caption[fractions]{Partial fractions in the SM using the form
factor model given in refs.\cite{Boito:2009qd, ElBennich:2009da}.
The constants $\chi_S$ and $\chi_P$ were adjusted to fit the total
BR=9.13(19) \% and the s-wave contribution \cite{pdg}. The
additional phase given in ref.\cite{Boito:2009qd} between the s
and p-waves was kept fixed to $\phi_{SP}=-65$. The SPI column is
the s and p-waves interference. The columns labelled I-IV
correspond to the contributions from the regions: I $s<0.7$
GeV$^2$, II $0.7\ {\rm GeV}^2<s<1$ GeV$^2$, III $1\ {\rm
GeV}^2<s<2.25$ GeV$^2$ and IV $s>2.25$ GeV$^2$. \label{fractions}}
\end{table}

\section{ CP asymmetry in $D^+ \to K^- \pi^+ \pi^+$ within SM  \label{CPSM}}

Non-vanishing direct CP asymmetry requires a weak CP violating
phase which is clearly absent at tree level in the process $D^+
\to K^- \pi^+ \pi^+$ as both $a_1$ and $a_2$ are real. As a
consequence one has to calculate the corrections to $a_1$ and
$a_2$ that can be generated at the loop level in order to generate
the weak CP violating phase. It turns out that the corrections are
very small as they are generated through box and di-penguin
diagrams\cite{Donoghue:1986cj,Petrov:1997fw,box}. Detailed
calculations of these corrections can be found in
ref.\cite{Delepine:2012xw}. The box contribution can lead to a
correction to the Wilson coefficient $c_2$ that can be written  as
\cite{box,He:2009rz,Delepine:2012xw}

\begin{eqnarray}
\Delta {c_2} =\frac{G_F\,m_W^2}{ \sqrt{2}\,\pi^2\,V_{cs}^*V_{ud}}
b_x
\end{eqnarray}
where

\begin{equation}
b_x \simeq 3.6\cdot 10^{-7} {\rm e}^{0.07\cdot i}
\end{equation}

The other corrections to the Wilson coefficients  are due to the
dipenguin diagrams  and are  given as
 \cite{Donoghue:1986cj,Petrov:1997fw,Chia:1983hd,Delepine:2012xw}

\begin{eqnarray}
\Delta a_1 &\simeq & 2.8\cdot 10^{-8}{\rm e}^{-0.004i} \nonumber \\
\Delta a_2 & \simeq & -2.0\cdot 10^{-9}{\rm e}^{0.07i}
\end{eqnarray}
 Clearly from these correction the predicted  direct CP asymmetry
is still so tiny roughly speaking of order $10^{-8}$ or even can
be smaller than that. This CP asymmetry is out of reach of current
experiments at LHCb and also of near future experiments such as
Super B factories at KEK. Thus this result motivates to extend
the study to include New Physics extensions of the SM as we will
consider in the next sections.
\section{New Physics \label{CPNP}}
Within New Physics possible complex couplings works as new sources
for the CP violating weak phase. Since the short range physics in
both  $D^0\to K^-\pi^+$ and $D^+\to K^-\pi^+\pi^+$ are the same we
expect that the New Physics models that enhance  the CP asymmetry
in $D^0\to K^-\pi^+$  will have the potential to enhance the CP
asymmetry in $D^+\to K^-\pi^+\pi^+$. In our earlier work on the CP
asymmetry of $D^0\to K^-\pi^+$  we have found that  the possible
candidates that enhanced the CP asymmetry are the models with
charged Higgs bosons and the Left Right models
\cite{Delepine:2012xw}. In these classes of models the Wilson
coefficients of the effective Hamiltonian governs the decay
process of our interest can receive contribution from tree level
diagrams and thus the complex phases in these Wilson coefficients
will be dominant as the complex phase in the SM are generated at
the loop level with large suppression as we showed in details in
ref.\cite{Delepine:2012xw}.

  In the next subsections, we study the direct CP asymmetry of
$D^+\to K^-\pi^+\pi^+$ in the framework of these two candidates of
New Physics beyond the SM. In addition we  consider a toy model
where we assume that only  $a_2$ acquires an extra weak phase of
20$^\circ$. We  use this toy model just to illustrate that one can
define several CP asymmetries and we show their behavior as a
function of the kinematic variables of the process.

\subsection{ CP asymmetry in $D^+ \to K^- \pi^+ \pi^+$ within  a
toy model}

\begin{table}
\begin{tabular}{|r|r|r|r|r|r|r|r|r|r|r|r|r|}   \hline
Model-fract.    &$\chi_S$& $\chi_P$ & $\phi_W$  & $f_i:\ S$ & $P$
&SPI &I & II & III & IV & Tot.  \\ \hline Toy mod. $f_i$. &4.9
&5.05 &20  &80.1 &17      &2.9         &5.1 &22.5 &63.9 &8.6  & \\
\hline $A_{\rm CP}$    &-    &-    & -  &-    &-       &-
&-0.5&-3.3 &-1.2 &26.6 &0.3 \\ \hline $f_i$           &4.91 &5.07
&-20 &80.2 &16.4    &3.4         &5.1 &24.2 &65.6 &5.1  & \\
\hline $A_{\rm CP}$    &     &-    &-   &-    & -      & &0.5 &
3.3 &1.1  &-26.4&-0.3 \\  \hline
\end{tabular}
\caption[fractions]{Partial fractions and $A_{\rm CP}$ in the toy
model, adding a CPV phase to $a_2$.  No pion-pion interaction was
considered, BR=9.13(19) \% \cite{pdg},  $\phi_{SP}=-65$.
\label{toyp}}
\end{table}

\begin{figure}
%\centering
\setlength{\unitlength}{1cm}
\begin{picture}(16,10)

\put(0,5.5){\includegraphics[scale=0.8]{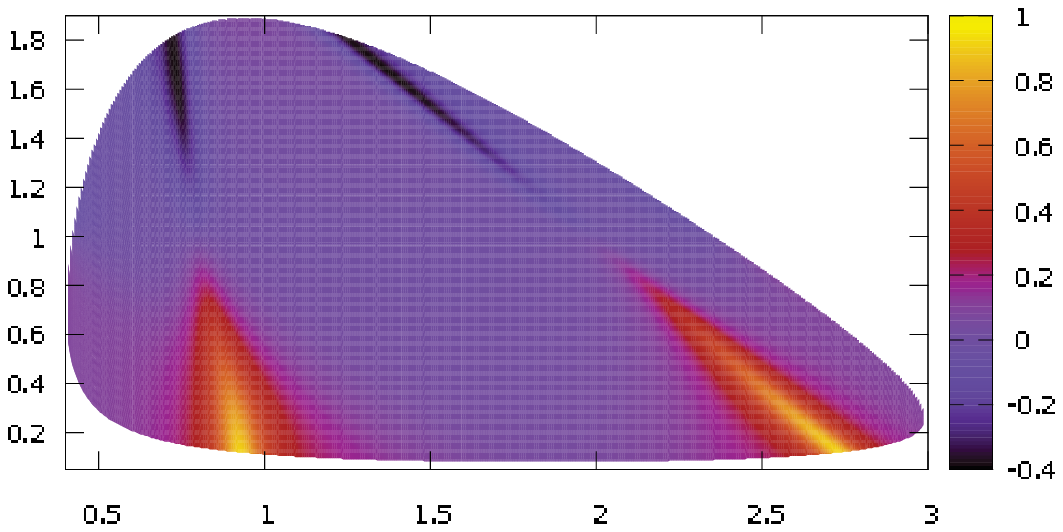}}
\put(-0.5,8){\rotatebox{90}{$u$  [GeV$^2$]}} \put(6.4,5){$s$
[GeV$^2$]} \put(6,9){\Large $A_{\rm CP}$}

\put(0,0){\includegraphics[scale=0.6]{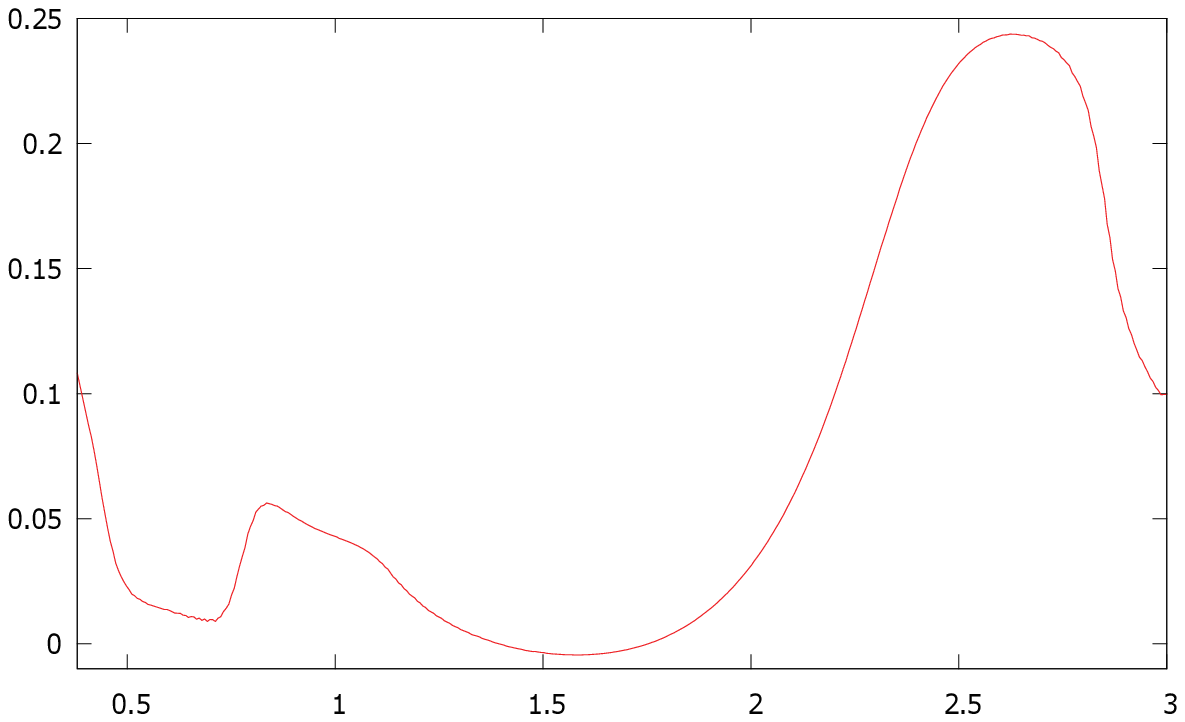}}
\put(-0.2,3.8){\rotatebox{90}{$A_{\rm CP}(s)$}} \put(5.7,0.7){$s$
[GeV$^2$]}

\put(8.5,5.4){\includegraphics[scale=0.6]{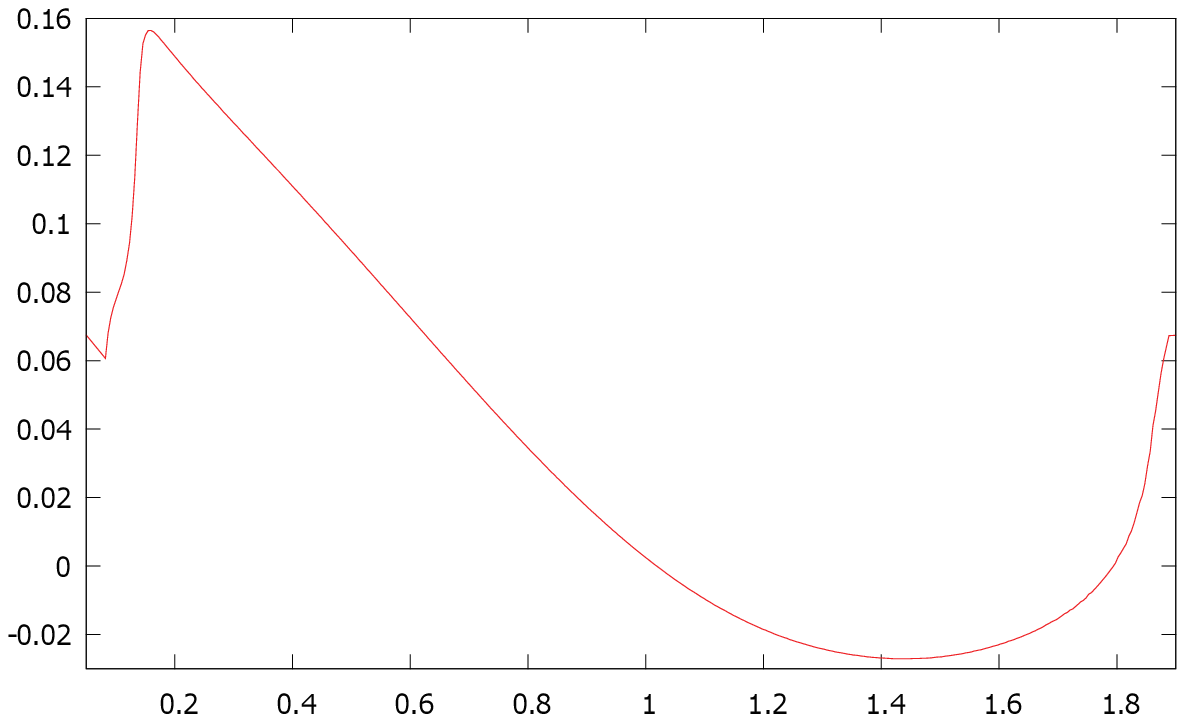}}
\put(15.2,8.5){\rotatebox{90}{\small $A_{\rm CP}(u)$}}
\put(14.4,5.2){$u$ [GeV$^2$]}
\end{picture}\vskip-0.5cm
\caption[$A_{\rm CP}$]{$A_{\rm CP}$ and its projections, for a toy
model where $a_2$ acquires an extra weak phase of
20$^\circ$.\label{toyf} }
\end{figure}

We consider a toy model where  $a_2$ acquires an extra weak phase
of 20$^\circ$.  Given that the two pions are identical one has to
average with the term where the two pions are interchanged: $\pi_1
\leftrightarrow \pi_2$, and $s\to t=s_0$. In this case one can
define several CPV asymmetries, like

\begin{eqnarray}
A_{\rm CP}(s=s_0,\ u) &=& {\left|A(s=s_0,u)\right|^2- \left|\bar
A(s=s_0,u)\right|^2 \over 2\left|A(s=s_0,u)\right|^2}+
{\left|A[t=s_0,u)]\right|^2- \left|\bar A(t=s_0,u)\right|^2 \over
2\left|A(t=s_0,u)\right|^2}
\nonumber \\
&=& {\left|A(s=s_0,u)\right|^2- \left|\bar A(s=s_0,u)\right|^2
\over \left|A(s=s_0,u)\right|^2}\simeq {2\sin\phi\over |A|^2}{\rm
Im} \left(A{A'}^*\right)
\nonumber \\
A_{\rm CP}(s=s_0) & =& {\int{\rm
d}u\left[\left|A(s=s_0,u)\right|^2-\left|\bar
A(s=s_0,u)\right|\right]^2 \over 2\int{\rm
d}u\left|A(s=s_0,u)\right|^2}+ {\int {\rm d}u
\left[\left|A[t=s_0,u)]\right|^2-\left|\bar
A(t=s_0,u)\right|^2\right] \over 2\int{\rm
d}u\left|A(t=s_0,u)\right|^2}
\nonumber \\
&=& {\int{\rm d}u \left[\left|A(s=s_0,u)\right|^2-\left|\bar
A(s=s_0,u)\right|^2\right] \over \int{\rm
d}u\left|A(s=s_0,u)\right|^2}  \simeq 2\sin\phi {\int {\rm
Im}\left(A{A'}^*\right)\ {\rm d}u\over \int |A|^2\ {\rm d}u}
\nonumber \\
A_{\rm CP,\ tot.} &=& {\int {\rm d}s{\rm d}u |{\bf p_K'}| |{\bf
p_2'}| \left[\left|A\right|^2-\left|\bar A\right|^2\right] \over
\int{\rm d}s{\rm d}u|{\bf p_K'}| |{\bf p_2'}| \left|A\right|^2}
\simeq 2\sin\phi {\int {\rm Im}\left(A{A'}^*\right)\ {\rm d}s{\rm
d}u\over \int |A|^2\ {\rm d}s{\rm d}u}
\end{eqnarray}

where $A$ is the dominant SM amplitude, $A'_{\rm NP}=A'{\rm
e}^{i\phi}$  is the New Physics   amplitude with $\phi$ its
CP-Violating(CPV)  phase \cite{RyanMackenzieWhite:2013qfa}. In
Fig.(\ref{toyf}) we show the plots of these asymmetries. In Table
\ref{toyp} we show the predicted partial fractions and the total
$A_{\rm CP}$. It is interesting to emphasize the fact that even if
the total CP asymmetry is small (around $0.003$, it is possible to
get  much larger CP asymmetries  restraining to one region of the
phase space parameters. For instance in this toy model, in region
IV, it is possible to get a CP asymmetry up to $25\%$.

\subsection{A new charged gauge boson as Left Right models}
In this section we consider  a well known candidate for NP beyond
the Standard Model based on extending the SM gauge group to
include a new  gauge group namely
$SU(2)_R$\cite{Pati:1973rp,Mohapatra:1974hk,Mohapatra:1974gc,
Senjanovic:1975rk,Senjanovic:1978ev}. Thus  our gauge group
defining the electroweak interaction is given by $SU(2)_L \times
SU(2)_R \times U(1)_{B-L}$. This extension of the SM  has been
widely studied in the literature (see for instance refs.
\cite{Beall:1981ze,Cocolicchio:1988ac,Langacker:1989xa,Cho:1993zb,Babu:1993hx}
) and the constraints on the parameter space of the model have
been derived using experimental measurements in refs.
\cite{Beringer:1900zz,Alexander:1997bv,Acosta:2002nu,Abazov:2006aj,Abazov:2008vj,Abazov:2011xs}.
 With the running of LHC,  CMS \cite{Chatrchyan:2012meb,Chatrchyan:2012sc} and
ATLAS \cite{Aad:2011yg,Aad:2012ej} collaborations have  improved
the bound on the scale of the $W_R$ gauge boson mass
\cite{Maiezza:2010ic}.

 There are two scenarios to be study in this context. The first
 one is to assume no mixing between $W_L$ and $W_R$ gauge
bosons while the second one is to allow mixing between these gauge
bosons.

 We start our analysis  with the first scenario. The new diagrams contributing
to $D^+ \to K^- \pi^+\pi^+$  are similar to the SM tree-level
diagrams with $W_L $ is replaced by a $W_R$. These diagrams lead
to new contributions to the effective Hamiltonian governs our
decay process that can be expressed as:

\begin{eqnarray}
{\cal H}^{NM}_{\rm LR} &=& {G_F\over\sqrt{2}}\left({g_R m_W\over
g_L m_{W_R}}\right)^2V_{Rcs}^*V_{Rud}\left(c_1'\bar s\gamma_\mu
c_R\bar u \gamma^\mu d_R+c_2'\bar u \gamma_\mu c_R\bar s\gamma^\mu
d_R\right)+{\rm h.c.}
\end{eqnarray}
where ${\cal H}^{NM}_{\rm LR}$ denotes  the effective Hamiltonian
in the case of no mixing,  $g_{R}$ and $g_{L}$ denote the gauge
$SU(2)_{R}$ and $SU(2)_{L}$ couplings respectively. The gauge
bosons associated with the gauge groups $SU(2)_{L}$ and
$SU(2)_{R}$ have masses $m_W$ and $m_{W_R}$ respectively. The
matrix $ V_R$ represents the quark mixing matrix that appears in
the right sector of the Lagrangian similar to the CKM quark mixing
matrix. The effective Hamiltonian ${\cal H}^{NM}_{\rm LR}$ lead to
a contribution to the amplitude of the decay process under
investigation. Although it is expected that the new contribution
to amplitude will  enhance the SM prediction for the CP asymmetry
but still it will be suppressed due to the limit on $M_{W_R}$
which has to be of order $2.3$ TeV in this case of no-mixing Left
right models \cite{Maiezza:2010ic}.

  We turn now to the second scenario where we assume mixing between
$W_L$ and $W_R$ gauge bosons. This scenario can strongly enhance
the CP violation in the Charm and muon sectors as has been
concluded in refs.\cite{Chen:2012usa,Lee:2011kn}. Recently we have
investigated this conclusion in the study of the CP asymmetry in
the decay channel $D^0\to K^-\pi^+$ where the enhanced asymmetry
was at the level of $10\%$ \cite{Delepine:2012xw}. This motivates
us to see what will be the maximum enhancement that can be reached
for the CP asymmetry in  $D^+ \to K^- \pi^+\pi^+$ in this
scenario.

  We start our investigation by relating the weak eigenstate, $W_L$ and
$W_R$, to the mass eigenstates,  $W_1$ and $W_2$ , of the
$SU(2)_{R}$ and $SU(2)_{L}$ gauge bosons via \cite{Chen:2012usa}
\begin{eqnarray}
\left(\begin{array}{c}
  W_L \\
  W_R
\end{array}\right) =
\left(\begin{array}{cc}
  \cos \xi & -\sin \xi \\
{\rm e}^{i\omega}\sin \xi & {\rm e}^{i\omega}\cos \xi
\end{array}\right)
\left(\begin{array}{c}
  W_1 \\
  W_2
\end{array}\right)\simeq
 \left(\begin{array}{cc}
  1 & - \xi \\
{\rm e}^{i\omega}\xi & {\rm e}^{i\omega}
\end{array}\right)
\left(\begin{array}{c}
  W_1 \\
  W_2
\end{array}\right)
\end{eqnarray}

where $\xi$ denotes the Left-Right (LR) mixing angle. Deviation to
the non-unitarity of the CKM quark mixing matrix can lead to
strong constraints on $\xi$ and on the right scale $M_R$. In  the
case that the couplings $g_R$ and $g_L$ are equal  at the
unification scale i.e. the Left-Right symmetry is  manifest,  the
mixing angle $\xi$ has to be smaller than
0.005\cite{Wolfenstein:1984ay} and the right scale $M_R$ has to be
bigger than 2.5 TeV\cite{Maiezza:2010ic}. As a consequence, one
expects that the predicted CP asymmetry will be so small also. On
the other hand and in the case where  $g_R$ is different than
$g_L$ at the unification scale i.e. the Left-Right symmetry is not
manifest, the limit on $M_R$ scale is much less restrictive and
the right gauge bosons could be as light as $0.3$ TeV
\cite{Olness:1984xb}. In this  case, $\xi$ can be as large as
$0.02$ if large CP violation phases in the right sector are
present \cite{Langacker:1989xa} still compatible with experimental
data \cite{Jang:2000rk,Badin:2007bv,Lee:2011kn}. It has shown
recently that, taking $g_L=g_R$,  the precision measurement of the
muon decay parameters done by TWIST collaboration
\cite{MacDonald:2008xf,TWIST:2011aa} can set model independent
limit on $\xi$ to be smaller than 0.03. We adopt this case in our
analysis and take  $\xi \sim 10^{-2}$.

\begin{table}
\begin{tabular}{|r|r|r|r|r|r|r|r|r|r|r|r|r|}   \hline
Model-fract.    &$\chi_S$& $\chi_P$ &$\phi_{LR}\ [^\circ]$& $f_i:\
S$ & $P$ &SPI &I &II &III & IV & Tot.  \\ \hline LR mod. $f_i$. &5
&5.655 &30  &80.2 &15.4  &4.3         &4.8 &23.7 &65.4 &6.1 & \\
\hline $A_{\rm CP}$    &-    &-     -  &-    &-       &- &-
&-0.4&1.2  &1.9  &-2.2 &1.3 \\ \hline

LR mod. $f_i$.  &4.956&5.64  &60  &80.2 &15.9  &3.9         &4.7
&23.7 &65.4 &6.1  & \\  \hline $A_{\rm CP}$    &-    &-     &- &-
&-     &-           &-0.6 &2.1  &3.2  &-3.7 &2.3 \\ \hline

LR mod. $f_i$.  &4.923&5.625 &90  &80.2&16.5 &3.3           &4.8
&23.7 &65.3 &6.2  & \\  \hline $A_{\rm CP}$    &-    &-     & - &-
&-    &-             &-0.7 &2.4  &3.7  &-4.1 &2.7 \\ \hline

LR mod. $f_i$.  &4.907&5.605 &120  &80.2 &17 &2.8           &4.9
&23.6 &65   &6.5  & \\  \hline $A_{\rm CP}$    &-    &-     & - &-
&-  &-             &-0.6 &2.1  &3.2  &-3.5 &2.3 \\ \hline

LR mod. $f_i$.  &4.913&5.59  &150  &80.2 &17.4&2.4          &5
&23.6 &64.7 &6.7  & \\  \hline $A_{\rm CP}$    &-    &-     & - &-
&-   &-            &-0.3 &1.2  &1.8  &-2   &1.3 \\ \hline

LR mod. $f_i$.  &4.974&5.59  &-150 &80.2 &17.4&2.5         &5.2
&23.5 &64.2 &7.1  & \\  \hline $A_{\rm CP}$    &-    &-     & - &-
&-   &-           &0.3   &-1.2 &-1.8 &2    &1.3 \\ \hline

LR mod. $f_i$.  &5.015&5.6   &-120 &80.2 &16.9&2.9         &5.2
&23.5 &64.2 &7.1  & \\  \hline $A_{\rm CP}$    &-    &-     & - &-
&-   &-           &0.5   &-2   &-3.1 &3.5 &-2.2 \\ \hline

LR mod. $f_i$.  &5.05 &5.62  &-90  &80.2 &16.4&3.4         &5.2
&23.5 &64.3 &7   & \\  \hline $A_{\rm CP}$    &-    &-     & - &-
&-   &-           &0.6   &-2.4 &-3.6 &4.1 &-2.6 \\ \hline

LR mod. $f_i$.  &5.067&5.64  &-60  &80.2 &15.8&4           &5.1
&23.6  &64.6 &6.8  & \\  \hline $A_{\rm CP}$    &-    &-     & -
&-    &-   &-           &0.5   &-2.1  &-3.2 &3.7  &-2.3 \\ \hline

LR mod. $f_i$.  &5.063&5.655 &-30  &80.2 &15.4&4.4         &5
&23.6 &64.9 &6.5  & \\  \hline $A_{\rm CP}$    &-    &-     & - &-
&-   &-           &0.3   &-1.2 &-1.8 &2.2  &-1.3 \\ \hline

\end{tabular}
\caption[fractions]{{ Partial fractions and $A_{\rm CP}$  with
$|c_{LR}|=0.02$ which corresponds to  its  maximum value allowed
from the Twist coll.\cite{MacDonald:2008xf, TWIST:2011aa}, for
different values of its phase. No pion-pion interaction was
considered, BR=9.13(19) \% \cite{pdg}, $\phi_{SP}=-65$.}
\label{LRCP}}
\end{table}

   The charged currents interaction can be written as

\begin{eqnarray}
{\cal L} &\simeq & -{1\over \sqrt{2}} \bar U \gamma_\mu
\left(g_LVP_L+g_R\xi \bar V^RP_R\right)DW_1^\dagger- {1\over
\sqrt{2}} \bar U \gamma_\mu \left(-g_L\xi VP_L+g_R\bar
V^RP_R\right)DW_2^\dagger
\end{eqnarray}

where $V=V_{\rm CKM}$ and $\bar V^R={\rm e}^{i\omega}V^R$.
Integrating out the $W_1$ in the usual way and neglecting the
$W_2$ contributions, given its mass is much higher, one obtains

\begin{eqnarray}
{\cal H}_{\rm eff.} &=& {4G_F\over \sqrt{2}}\left[c_1\bar
s\gamma_\mu \left(V^*P_L+{g_R\over g_L}\xi \bar V^{R*}P_R
\right)_{cs}c\bar u\gamma^\mu  \left(VP_L+{g_R\over g_L}\xi \bar
V^RP_R  \right)_{ud}  d \right.
\nonumber  \\
&& +\left. c_2\bar s_\alpha \gamma_\mu \left(V^*P_L+{g_R\over
g_L}\xi \bar V^{R*}P_R  \right)_{cs}c_\beta\bar u_\beta\gamma^\mu
\left(VP_L+{g_R\over g_L}\xi \bar V^RP_R  \right)_{ud} d_\alpha
\right]+{\rm h.\ c.}
\end{eqnarray}

The  terms proportional to $\xi$ of the effective Hamiltonian lead
to
\begin{eqnarray}
\Delta{\cal H}_{\rm eff.} &\simeq & {G_F\over \sqrt{2}}{g_R\over
g_L}\xi \left[ V_{cs}^* \bar V^R_{ud} \left( c_1\bar s\gamma_\mu
c_L\bar u \gamma^\mu d_R -2c_2\bar uc_L \bar s d_R  \right)+\bar
V^{R*}_{cs}V_{ud} \left(c_1\bar s\gamma_\mu c_R \bar u \gamma^\mu
d_L-2c_2 \bar u c_R \bar s d_L\right) \right] +{\rm h.c.}\nonumber\\
\end{eqnarray}
It is direct, see Appendix \ref{FFD} for matrix elements of the
operators, to show that $\Delta{\cal H}_{\rm eff.}$ result in a
new contribution to the amplitude of $D^+\to K^-\pi^+\pi^+$ given
by
\begin{eqnarray}
\delta A_{D^+\to K^-\pi^+\pi^+}&=& -{G_F\over \sqrt{2}}{g_R\over
g_L}\xi \left(V_{cs}^*\bar V^R_{ud}+\bar V^{R*}_{cs}
V_{ud}\right)\left( a_1 A_1+ 2a_2M_{D\to K
\pi\pi}^2F_0^{K^-\pi^+}(s)F_0^{D^+\pi^+}(s)  \right)\nonumber\\
\end{eqnarray}
 and thus the total amplitude becomes
\begin{eqnarray}
A_{D^+\to K^-\pi^+\pi^+}&=& {G_F\over
\sqrt{2}}V_{cs}^*V_{ud}\left[a_1(1-c_{LR})A_1+a_2[A_2-2c_{LR}M_{D\to
K \pi\pi}^2F_0^{K^-\pi^+}(s)F_0^{D^+\pi^+}(s)]\right]\nonumber\\
\end{eqnarray}

with $c_{LR}=(g_R\xi/g_L)\left(\bar V^R_{ud}/V_{ud}+\bar
V^{R*}_{cs}/V_{cs}^*\right)$.
 We start our numerical analysis for  the CP asymmetry
 by  parameterizing $c_{LR}$ as $c_{LR}=|c_{LR}|{\rm e}^{i\phi_{LR}}$.  In Table
\ref{LRCP} we show the corresponding predictions for the  partial
fractions and CP asymmetry as a function of the phase $\phi_{LR}$
for  $|c_{LR}|=0.02$, the maximum value allowed from the Twist
coll.\cite{MacDonald:2008xf, TWIST:2011aa}, for different values
of the  phase $\phi_{LR}$. We see from the table that the partial
fractions and the predicted CP asymmetry varies with the phase
$\phi_{LR}$ as expected but can be as large as $25\%$ in some
cinematical region of phase space. Thus measuring a large CP
asymmetry in this decay channel would be a hint of LR symmetric
model with mixing between $W_L$ and $W_R$ gauge bosons.
\section{Models with Charged Higgs contributions}\label{SMHeff}
Charged Higgs appears in extensions of the Higgs sector of the SM
as one of the physical mass eigenstates.  In one of these
extensions a new $SU(2)_L$ doublet is added to the Higgs sector of
the SM.  With this new doublet, there are several possibilities to
couple the two Higgs doublets to the fermions. As a consequence we
have different types of these two Higgs doublet models (2HDM) such
as type I, II or III  and so on (for a review see ref.
\cite{Branco:2011iw}).  LEP has performed a Direct search for a
charged Higgs in 2HDM type II.  They obtained a bound on the
charged Higgs mass of $78.6$ GeV \cite{Searches:2001ac}. Recently
the  results on $ B\to \tau \nu$ obtained by  BABAR
\cite{Aubert:2009wt}  and  BELLE \cite{Hara:2010dk} have improved
the indirect constraints on the charged Higgs mass in type II 2HDM
\cite{Baak:2011ze}:
\begin{equation}
m_{H^+}> 240 GeV \ \  at \ \ 95 \% CL
\end{equation}
 In our study we will adopt 2HDM type III which
is a general model where both two Higgs doublets can couple to up
and down quarks. This means that 2HDM type III can lead to Flavor
changing neutral currents and thus  they can be used to strongly
constrain the new parameters in the model. With the presence of
the complex couplings in the model that escape the strong
constraints one expects to have a sizable contribution to the
direct CP asymmetry as we have shown in our earlier work on  the
CP asymmetry of  $D^0\to K^-\pi^+$\cite{Delepine:2012xw}.

The Yukawa Lagrangian of the 2HDM of type III is given as
\cite{Crivellin:2010er,Crivellin:2012ye}:
\begin{eqnarray}
\mathcal{L}^{eff}_Y &=& \bar{Q}^a_{f\,L} \left[
  Y^{d}_{fi} \epsilon_{ab}H^{b\star}_d\,-\,\epsilon^{d}_{fi} H^{a}_u \right]d_{i\,R}\\
&-&\bar{Q}^a_{f\,L} \left[ Y^{u}_{fi}
 \epsilon_{ab} H^{b\star}_u \,+\, \epsilon^{ u}_{fi} H^{a}_d
  \right]u_{i\,R}\,+\,\rm{H.c}. \,,\nonumber
\end{eqnarray}
here $\epsilon_{ab}$ is the totally antisymmetric tensor, and
$\epsilon^q_{ij}$ stands for  the non-holomorphic corrections that
couple up (down) quarks to the down (up) type Higgs doublet. After
electroweak symmetry breaking, the effective Lagrangian
$\mathcal{L}^{eff}_Y$ gives rise to the  Higgs couplings to quarks
given as:
\begin{equation}
i\left({\Gamma_{u_f d_i }^{H^\pm\,LR\,\rm{eff} } }P_R+{\Gamma_{u_f d_i }^{H^\pm\,RL\,\rm{eff} } }P_L\right)\, ,\\
 \label{Higgs-vertex}
\end{equation}
with
\begin{eqnarray}
{\Gamma_{u_f d_i }^{H^\pm\,LR\,\rm{eff} } } &=& \sum\limits_{j =
1}^3 {\sin\beta\, V_{fj} \left( \frac{m_{d_i }}{v_d} \delta_{ji}-
  \epsilon^{ d}_{ji}\tan\beta \right), }
\\
{\Gamma_{u_f d_i }^{H^ \pm\,RL\,\rm{eff} } } &=& \sum\limits_{j =
1}^3 {\cos\beta\,  \left( \frac{m_{u_f }}{v_u} \delta_{jf}-
  \epsilon^{ u\star}_{jf}\tan\beta \right)V_{ji}}\,.\nonumber
 \label{Higgsv}
\end{eqnarray}

where $v_d$ and $v_u$  are the vacuum expectations values of  the
neutral component of the  Higgs doublets and  $V$ denotes  the CKM
matrix. The Feynman-rule given in Eq. (\ref{Higgs-vertex}) can
lead to the effective Hamiltonian  that governs the process under
consideration after integrating out the charged Higgs mediating
the tree diagram

 \be {\mathcal H}^{H\pm}_{eff}= \frac{
G_F}{\sqrt{2}}V^*_{cs}V_{ud} \sum^4_{i=1} C^H_i(\mu)
Q^H_i(\mu),\ee  here $C^H_i$ denotes the Wilson coefficients
obtained by perturbative QCD running from $M_{H^{\pm}}$ scale to
the scale $\mu$ relevant for hadronic decay and $Q^H_i$ are the
corresponding local operators at low energy scale
$\mu\simeq m_c$. These operators can be written as %
\bea
Q^H_1 &=&(\bar{s} P_R c)(\bar{u} P_L d),\nonumber\\
Q^H_2 &=&(\bar{s} P_L c)(\bar{u} P_R d),\nonumber\\
Q^H_3 &=&(\bar{s} P_L c)(\bar{u} P_L d),\nonumber\\
Q^H_4 &=&(\bar{s} P_R c)(\bar{u} P_R d), \label{operator} \eea and
their corresponding  Wilson coefficients $C^H_i$, at the
electroweak scale, are given by
\begin{eqnarray}
C^H_1 &=& \frac {\sqrt{2} }{ G_F V^*_{cs}V_{ud}
 m^2_H} \bigg(\sum\limits_{j = 1}^3
{\cos\beta\, V_{j1} \left( \frac{m_u }{v_u} \delta_{j1}-
\epsilon^{ u\star}_{j1}\tan\beta \right)}\bigg)\bigg(
\sum\limits_{k= 1}^3 {\cos\beta\,V^{\star}_{k2}} \left(
\frac{m_c}{v_u} \delta_{k2}-\epsilon^{ u}_{k2}\tan\beta
\right)\bigg),\nonumber\\
C^H_2 &=& \frac {\sqrt{2} }{ G_F V^*_{cs}V_{ud}
 m^2_H} \bigg(\sum\limits_{j = 1}^3
{\sin\beta\,V_{1j}  \left( \frac{m_d }{v_d} \delta_{j1}-
\epsilon^{ d}_{j1}\tan\beta \right)}\bigg)\bigg( \sum\limits_{k=
1}^3 {\sin\beta\,V^{\star}_{2k}} \left( \frac{m_s}{v_d}
\delta_{k2}-\epsilon^{ d\star}_{k2}\tan\beta
\right)\bigg)\nonumber\\
 C^H_3 &=& \frac {\sqrt{2} }{  G_F V^*_{cs}V_{ud}
 m^2_H} \bigg(\sum\limits_{j = 1}^3 {\cos\beta\, V_{j1} \left(
\frac{m_u }{v_u} \delta_{j1}- \epsilon^{ u\star}_{j1}\tan\beta
\right)}\bigg)\bigg( \sum\limits_{k= 1}^3
{\sin\beta\,V^{\star}_{2k}} \left( \frac{m_s}{v_d}
\delta_{k2}-\epsilon^{ d\star}_{k2}\tan\beta
\right)\bigg),\nonumber\\
C^H_4 &=& \frac {\sqrt{2} }{ G_F V^*_{cs}V_{ud}
 m^2_H}\bigg( \sum\limits_{k= 1}^3
{\cos\beta\,V^{\star}_{k2}} \left( \frac{m_c}{v_u}
\delta_{k2}-\epsilon^{ u}_{k2}\tan\beta
\right)\bigg)\bigg(\sum\limits_{j = 1}^3 {\sin\beta\,V_{1j} \left(
\frac{m_d }{v_d} \delta_{j1}- \epsilon^{ d}_{j1}\tan\beta
\right)}\bigg) \nonumber \\
 \label{Higgscoe}
\end{eqnarray}

The  contribution of the charged Higgs to the amplitude of the
decay process under consideration can be obtained via
  \begin{eqnarray}
<K^-\pi^+\pi^+|{\mathcal H}^{H\pm}_{eff}|D^+>\, \equiv \,\delta
A^{H^{\pm}}_{D^+\to K^-\pi^+\pi^+}\end{eqnarray}

To calculate the matrix element in the last equation we first use
Fierz's identities to rewrite the set of the operators in
Eq.(\ref{operator}) in a new basis include only vector, axial
vectors and tensor operators. Second we can easily write the
vector and axial vector operators in terms of $A_1$ and $A_2$ and
for the matrix element of the tensor operator we  can parametrize
it as we will show below. Thus we find
\begin{eqnarray}
<K^-\pi^+\pi^+|\bar s\,P_L\, c\, \bar u \,P_R\, d |D^+>
&=&<K^-\pi^+\pi^+|\bar s\, \,P_R \,c \,\bar u \,P_L \,d|D^+> \nonumber \\
&=&{m_\pi^2\over (m_c+m_s)(m_u+m_d)} A_1-{1\over 2N}A_2
\nonumber \\
<K^-\pi^+\pi^+|\bar s \,P_L\, c\, \bar u \,P_L\, d|D^+>
&=&<K^-\pi^+\pi^+|\bar s \,P_R\, c\, \bar u \,P_R \,d|D^+>
=-{m_\pi^2\over (m_c+m_s)(m_u+m_d)} A_1\nonumber \\
&+&{1\over 2N}{\Delta_{D\pi}^2\Delta_{K\pi}^2 \over
(m_c-m_u)(m_s-m_d)}F_0^{K^-\pi^+}(s)F_0^{D^+\pi^+}(s)+T
\nonumber \\
\end{eqnarray}
here $T$ represents the contribution of the tensor operators.  The
explicit form of $T$  can be obtained upon calculating the matrix
elements of the tensor operators that can written as
\begin{eqnarray}
<K^-\pi^+\pi^+|\bar s\sigma^{\mu\nu}\,P_L\,d \,\bar u\,
\sigma_{\mu\nu}\, P_L \, c |D^+> &=&  <K^-\pi^+_1|\bar
s\sigma^{\mu\nu} \,P_L\, d |0> < \pi^+_2| \bar u \sigma_{\mu\nu}
\,P_L\, c |D^+>\nonumber\\<K^-\pi^+\pi^+|\bar
s\sigma^{\mu\nu}\,P_R\,d \,\bar u\, \sigma_{\mu\nu}\, P_R  c |D^+>
&=&  <K^-\pi^+_1|\bar s\sigma^{\mu\nu} \,P_R\, d |0> < \pi^+_2|
\bar u \sigma_{\mu\nu} \,P_R c |D^+>\label{Tens1}
\,\,\end{eqnarray}

Using the kinematic of the decay process it is direct to
parameterize the matrix elements in the last equation as (see
Appendix \ref{FFD} for details)
\begin{eqnarray}
<K^-\pi^+_1\pi^+_2|\bar s\,\sigma^{\mu\nu}\,P_L\, d \,\bar u
\sigma_{\mu\nu} \,P_L \,c |D^+> &=& <K^-\pi^+_1\pi^+_2|\bar
s\,\sigma^{\mu\nu} \,P_R\, d\, \bar u\, \sigma_{\mu\nu} \,P_R \,c
|D^+> \nonumber
\\&=& 4\, h^{K^-\pi^+}(s)\, h^{D^+\pi^+}(s) \bigg[s(t-u)-\Delta^2_{D\pi}\Delta^2_{K\pi}\bigg]\label{tens3}
\end{eqnarray}
Thus finally we get
\begin{eqnarray}
\delta A^{H^{\pm}}_{D^+\to K^-\pi^+\pi^+}&=&
(C^H_1+C^H_2-C^H_3-C^H_4)\chi^{\pi^+}A_1-\frac{1}{2 N}
(C^H_1+C^H_2)A_2\nonumber \\&+&\frac{1}{2 N}(C^H_3+C^H_4)
{\Delta_{D\pi}^2\Delta_{K\pi}^2 \over
(m_c-m_u)(m_s-m_d)}F_0^{K^-\pi^+}(s)F_0^{D^+\pi^+}(s)\nonumber \\
&+&\frac{1}{2 N}(C^H_3+C^H_4)
\bigg[s(t-u)-\Delta^2_{D\pi}\Delta^2_{K\pi}
\bigg]h^{D^+\pi^+}(s)\,
h^{K^-\pi^+}(s)\label{Higgscont}\end{eqnarray} where
 \begin{eqnarray} \chi^{\pi^+}={m_\pi^2\over
(m_c+m_s)(m_u+m_d)}\end{eqnarray} and we can write the total
amplitude as
\begin{eqnarray} A_{D^+\to
K^-\pi^+\pi^+}&=& {G_F\over
\sqrt{2}}V_{cs}V_{ud}\left(a_1A_1+a_2A_2\right) +\delta
A^{H^{\pm}}_{D^+\to K^-\pi^+\pi^+}
\end{eqnarray}
 We now discuss the experimental constraints on the parameters
$ \epsilon^{ q}_{ij}$ where $q=d,u$ that appear in the Wilson
coefficients. In the down sector, $ \epsilon^{d}_{ij}$,  we find
that for $i\neq j$ the parameters $\epsilon^d_{ij}$ are strongly
constrained from FCNC processes in the down sector because of
tree-level neutral Higgs exchange. Hence, we are left only with
$\epsilon^d_{11}$ and $\epsilon^d_{22}$.  On the other hand in the
up sector,  $\epsilon^u_{ij}$,   we  note  that only the terms
proportional to $\epsilon^u_{11}$ and $\epsilon^u_{22}$ can
significantly affect the Wilson coefficients without any CKM
suppression factors. Other $\epsilon^u_{ij}$ terms will be
suppressed by orders $\lambda$ or $\lambda^2$ or higher and so we
can safely neglect them in the analysis.

  The naturalness criterion of 't Hooft set sever constraints on
the $\epsilon^d_{11}$, $\epsilon^d_{22}$ and $\epsilon^u_{11}$ due
to the smallness of the down, strange and up quark masses
respectively  while $\epsilon^{ u}_{22}$ become less constrained
as discussed in Refs.\cite{Delepine:2012xw,Crivellin:2013wna}.
Thus we keep in our analysis terms that proportional to
$\epsilon^{ u}_{22}$ and drop the terms that are proportional to
$\epsilon^d_{11}$, $\epsilon^d_{22}$ and $\epsilon^u_{11}$. The
relevant constraints on $\epsilon^{ u}_{22}$ have been discussed
in details in Refs.\cite{Delepine:2012xw,Crivellin:2013wna} and
thus we take into account these constraints in the analysis of the
CP asymmetry below.

 We start our analysis for Higgs contribution to the CP asymmetry
 by  parameterizing $\epsilon^{ u}_{22}$ as $\epsilon^{
u}_{22}=|\epsilon^{ u}_{22}|{\rm e}^{i\phi_W}$. From the
constraints discussed above, as an example,  we can take
$|\epsilon^{ u}_{22}|=0.7$ for $m_H=240 \,GeV$ and
$\tan\beta=100$. In Table \ref{Higs} we show the corresponding
predictions for the  partial fractions and CP asymmetry as a
function of the phase $\phi_W$. We see from the table that the
partial fractions are almost constant to the value in the first
row. On the other hand, from  Table \ref{Higs},  we note  that the
predicted CP asymmetry varies with the phase $\phi_W$ and can
reach a maximum value of $A_{CP}\simeq -1.4\times 10^{-3}$. For
larger values of $m_H$ and smaller values of $\tan\beta$ we find
that the predicted asymmetry $A_{CP} \leq{\cal O} (10^{-4})$.

\begin{table}\centering
\begin{tabular}{|r|r|r|r|r|r|r|r|r|r|r|r|r|}   \hline
Model-fract.    &$\chi_S$& $\chi_P$ & $\phi_W$  & $f_i:\ S$ & $P$
&SPI &I & II & III & IV &  \\ \hline Higgs-frac      &4.96
&5.565     & 30  &80.2 &16.4  & 3.3  &5 &23.4 &65 &6.5&  \\ \hline
Higgs-$10^4\times A_{\rm CP}$ &-    &-         & -   &-    &-
& -    &1.4 &-1.6 &-8.2 &-19 & -6.9 \\ \hline Higgs-$10^4\times
A_{\rm CP}$ &4.975&5.58      & 60  &-    &-     & -    &2.5 &-2.8
&-14  &-33& -12 \\ \hline Higgs-$10^4\times A_{\rm CP}$ &4.994&5.6
& 90  &-    &-     & -    &2.9 &-3.2 &-17  &-39 & -14 \\ \hline
Higgs-$10^4\times A_{\rm CP}$ &5.01 &5.625     & 120 &-    &-
& -    &2.6 &-2.8 &-14  &-34 & -12 \\ \hline Higgs-$10^4\times
A_{\rm CP}$ &5.025&5.64      & 150 &-    &-     & -    &1.5 &-1.6
&-8.3 &-20 & -7 \\ \hline
\end{tabular}
\caption[fractions]{Partial fractions and $A_{\rm CP}$. In the
Higgs case with  $|\epsilon^{ u}_{22}|=0.7$ , $\tan\beta=100$ and
$m_H=240$ GeV. The BR=9.13(19) \% \cite{pdg}, $\phi_{SP}=-65$. The
partial fractions are almost constant to the value in the first
row.}\label{Higs}
\end{table}

\section{Conclusion}

In this paper, we have studied the Cabibbo favored non-leptonic
$D^+\to K^- \pi^+ \pi^+$ decay.  We have shown that the direct CP
asymmetry in this decay mode within SM is strongly suppressed and
out of experimental range. Then we have explored new physics
models namely, a toy model with CP violating weak phase equals
$20^{\circ}$ in $a_2$, a model with extra gauge bosons within
Left-Right Grand Unification models and a model with charged Higgs
Field.

 The toy model strongly improved SM prediction of the CP
asymmetry where  the predicted CP asymmetry can reach $30\%$. This
asymmetry is large and if confirmed it will be an indication of NP
beyond SM and it will be challenging to find a New Physics
extension of the SM that can produce this weak phase in $a_2$
only.

  The next model which is most promising is non-manifest
Left-Right extension of the SM where the left right mixing between
the gauge bosons leads to a strong enhancement in the CP
asymmetry. In this class of  models, it is possible to get large
CP asymmetry $25\%$ which can be tested in the LHCb and the next
generation of charm or B factories.

  Our last model, the 2HDM type III, can lead to a
CP asymmetry that  depends on the charged Higgs masses and
couplings. A maximal value approximately a maximum value of
$A_{CP}\simeq -1.4\times 10^{-3}$ can be reached with a Higgs mass
of $240$ GeV and large tan$\beta$. Larger values of charged Higgs
mass lead to a smaller direct CP asymmetries.

 \section*{Acknowledgements}
 D. D. is  grateful to
Conacyt (M\'exico) S.N.I. and Conacyt project (CB-156618), DAIP
project (Guanajuato University) and PIFI (Secretaria de Educacion
Publica, M\'exico) for financial support. G. Faisel work is
supported by research grants NTU-ERP-102R7701 (National Taiwan
University).

\section*{ Appendix \label{FFD}}

For  $D^+\to K^-\pi^+\pi^+$ the expectation values of the
corresponding Left Right operators are

\begin{eqnarray}
<K^-\pi^+\pi^+|\bar s\gamma_\mu c_L \bar u \gamma^\mu d_R |D^+>
&=& <K^-\pi^+\pi^+|\bar s\gamma_\mu c_R \bar u \gamma^\mu d_L|D^+>\nonumber \\
& =& -A_1-{2\over N}M_{D\to K
\pi\pi}^2F_0^{K^-\pi^+}(s)F_0^{D^+\pi^+}(s)
\nonumber \\
<K^-\pi^+\pi^+|\bar s d_R\bar u c_L|D^+>&=&<K^-\pi^+\pi^+|\bar s
d_L \bar u c_R|D^+>\nonumber \\
& =& {1\over 2N}A_1+M_{D\to K
\pi\pi}^2F_0^{K^-\pi^+}(s)F_0^{D^+\pi^+}(s)
\end{eqnarray}
where the expression for $A_1$ can be found in
ref.\cite{Boito:2009qd} and
\begin{eqnarray}
M_{D\to K \pi\pi}^2 = {\Delta_{D\pi}^2\Delta_{K\pi}^2\over
(m_c-m_u)(m_s-m_d)}  \label{evLR}
\end{eqnarray}

In the case of the charged Higgs one has

\begin{eqnarray}
&&<K^-\pi^+\pi^+|\bar sc_L \bar u d_R |D^+> =<K^-\pi^+\pi^+|\bar s
c_R \bar u d_L|D^+> ={m_\pi^2\over (m_c+m_s)(m_u+m_d)} A_1-{1\over
2N}A_2
\nonumber \\
&&<K^-\pi^+\pi^+|\bar sc_L \bar u d_L |D^+> =<K^-\pi^+\pi^+|\bar s
c_R \bar u d_R|D^+>
\nonumber \\
&&=-{m_\pi^2\over (m_c+m_s)(m_u+m_d)} A_1+{1\over
2N}{\Delta_{D\pi}^2\Delta_{K\pi}^2 \over
(m_c-m_u)(m_s-m_d)}F_0^{K^-\pi^+}(s)F_0^{D^+\pi^+}(s)
\nonumber \\
&&+{1\over
2N}\left[s(t-u)-\Delta_{D\pi}^2\Delta_{K\pi}^2\right]h^{K^-\pi^+}(s)h^{D^+\pi^+}(s)
\end{eqnarray}

where the $h$ form factors from the tensor part and  the
expression for $A_1$ and  $A_2$ can be found in
ref.\cite{Boito:2009qd}.

\end{document}